\documentclass[12pt]{article}
\usepackage{epsfig, amssymb}
\usepackage{amsmath}
\usepackage{graphicx,epsfig}
\usepackage{color} 
\usepackage{cite}
\usepackage{hyperref}
\usepackage[title]{appendix}
\begin{document}
\thispagestyle{empty}
\begin{flushright}
\end{flushright}

\bigskip

\begin{center}
\noindent{\Large \textbf
{Stochastic quantization and holographic Wilsonian renormalization group of conformally coupled scalar in AdS$_{4}$
}}\\ 
\vspace{2cm} \noindent{Jun Hyeok Lee${}^{a}$\footnote{e-mail:junseyu2@gmail.com } and Jae-Hyuk Oh${}^{a}$\footnote{e-mail:jaehyukoh@hanyang.ac.kr}}

\vspace{1cm}
  {\it
Department of Physics, Hanyang University, Seoul 04763, Korea${}^{a}$\\
 }
\end{center}

\vspace{0.3cm}
\begin{abstract}
\noindent
In this paper, we explore the relationship between holographic Wilsonian renormalization groups and stochastic quantization in conformally coupled scalar theory in AdS$_{4}$. The relationship between these two different frameworks is firstly proposed in \href{https://arxiv.org/abs/1209.2242}{arXiv:1209.2242} and tested in various free theories. However, research on the theory with interactions has recently begun. In this paper, we show that the stochastic four-point function obtained by the Langevin equation is completely captured by the holographic quadruple trace deformation when the Euclidean action $S_{E}$ is given by $S_{E}=-2I_{os}$ where $I_{os}$ is the holographic on-shell  action in the conformally coupled scalar theory in AdS$_{4},$ together with a condition that the stochastic fictitious time $t$ is also identified with AdS radial variable $r$. We extensively explore a case that the boundary condition on the conformal boundary is Dirichlet boundary condition, and in that case, the stochastic three-point function trivially vanishes. This agrees with that the holographic triple trace deformation vanishes when Dirichlet boundary condition is applied on the conformal boundary.

\end{abstract}
\newpage

\section{Introduction}
\label{introduction}
AdS/CFT correspondence\cite{Aharony:1999ti,Maldacena:1997re} claims that quantum theory of gravity in $d+1$-dimensional anti-de Sitter space (AdS) corresponds to $d$-dimensional conformal field theory (CFT) on the AdS boundary. In AdS/CFT correspondence, the radial coordinate ``$r$'' defined in AdS where the quantum theory of gravity lives corresponds to the energy scale of the conformal field theory on AdS boundary\cite{Susskind:1998dq}. That is, CFT on $r=0$ boundary is defined in high energy scale, while the CFT on the $r=\infty$ boundary is defined in low energy scale. The radial flow of the gravity solutions can be interpreted as the flow of the renormalized group of boundary field theory. This interpretation is called a holographic renormalization group and is of great help in studying systems with strongly coupled fields\cite{Faulkner:2010jy,Faulkner:2010tq}.

The equation governing the radial flow of the quantum theory of gravity is the holographic Hamilton-Jacobi equation:

\begin{equation}
\label{1.1}
\partial_{\epsilon}S_B(\epsilon,\phi)=\int_{r= \epsilon}d^3p\left (\mathcal{L}-\cfrac{\partial S_B}{\partial\phi_{p}}\cfrac{\partial S_B}{\partial\phi_{-p}}\right) = \int_{r= \epsilon}d^3p \mathcal{H}_{RG}\left (\cfrac{\partial S_B}{\partial\phi_{p}},\phi\right),
\end{equation}

\noindent where $\epsilon$ is an arbitrary cut-off in the radial direction of ``$r$'', $\mathcal{L}$ is the Lagrangian density of the gravity theory defined in AdS$_{d+1}$, $\mathcal{H}_{RG}$ is the Hamiltonian of the holographic renormalization group obtained by Legendre transformation of $\mathcal{L}$, and $\mathcal{S_B}$ is the boundary deformation on $r=\epsilon$ boundary. This Hamilton-Jacobi equation (\ref{1.1}) is a semi-classical approximation of a Schrödinger type equation, which will be explained in the following. The Schrödinger type equation is

\begin{equation}
\label{1.3}
\partial_{\epsilon}\psi_H(\phi,t)=-\int_{r=\epsilon}d^3p\mathcal{H}_{RG}\left (\cfrac{\delta}{\delta\phi},\phi\right)\psi_H(\phi,t),
\end{equation}

\noindent where we define the wave function $\psi_H$ as

\begin{equation}
\psi_H(\phi,t)=e^{-S_B(\epsilon)}.
\end{equation}

\noindent When we take a semi-classical approximation as $\left(\cfrac{\delta S_B}{\delta\phi_{p}}\right)^2>>\cfrac{\delta^2 S_B}{\delta{\phi_{p}}^2}$, then the equation (\ref{1.3}) becomes (\ref{1.1}).

It's interesting that there are some papers \cite{Lifschytz:2000bj,Polyakov:2000xc,Mansi:2009mz} where the authors discuss  that AdS/CFT can be understood in the context of stochastic quantization. Stochastic quantization, as argued by Parisi \& Wu in \cite{Parisi:1980ys}, is a quantization method for Euclidean field theories. In this framework, fields are coupled with Gaussian noise of which randomness causes stochastic processes in the field. This system evolves along an imaginary time ``$t$''. The stochastic correlation function is given by

\begin{equation}
\label{1.4}
<\phi_p(t)\cdots\phi_{p'}(t)>=\int\mathcal{D}\phi P(\phi,t)(\phi_p(t)\cdots\phi_{p'}(t)),
\end{equation}

\noindent $P(\phi,t)$ is a stochastic probability distribution. After a long time $t\rightarrow\infty$, the correlation function (\ref{1.4}) approaches correlation function of the Euclidean quantum field theory. 

The equation governing the stochastic process is the Langevin equation, which is given by

\begin{equation}
\label{Lag-ge-bin}
\frac{\partial \phi(x,t)}{\partial t}=-\frac{1}{2}\frac{\delta S_E}{\delta\phi(x,t)}+\eta(x,t),
\end{equation}

\noindent where $\phi(x,t)$ is the stochastic field, ${S_E}$ is the field action, and $\eta(x,t)$ is Gaussian noise. From the Langevin equation, an equation for the stochastic probability distribution can be derived, which is called the Fokker-Planck equation,

\begin{equation}
\frac{\partial P(\phi,t)}{\partial t}=\frac{1}{2}\int d^dx \frac{\delta}{\delta \phi (x,t)} \left (\frac{\delta S_E}{\delta \phi (x,t)}+\frac{\delta}{\delta \phi (x,t)}\right) P(\phi,t).
\end{equation}

\noindent Stochastic quantization is also described by Schrödinger-type equations. To see this we define a wave function as follows:

\begin{equation}
\psi_S(\phi,t)=P(\phi,t)e^{\frac{S_E}{2}}.
\end{equation}

\noindent Then, one can easily show that the wave function satisfies the Schrödinger type equation,

\begin{equation}
\label{1.7}
\partial_{t}\psi_S(\phi,t)=-\int d^dx\mathcal{H}_{FP}\left (\frac{\delta}{\delta\phi},\phi\right)\psi_S(\phi,t),
\end{equation}

\noindent where $H_{FP}$ is the Fokker-Planck Hamiltonian,

\begin{align}
\mathcal{H}_{FP}=&\frac{1}{2}\left(-\frac{\delta}{\delta\phi(x)}+\frac{1}{2}\frac{\delta S_E}{\delta \phi(x)}\right)\left(\frac{\delta}{\delta\phi(x)}+\frac{1}{2}\frac{\delta S_E}{\delta \phi(x)}\right)
\\ \nonumber
=&-\frac{1}{2}\frac{\delta^2}{\delta\phi^2(x)}+\frac{1}{8}\left (\frac{\delta S_E}{\phi(x)}\right)^2-\frac{1}{4}\frac{\delta^2 S_E}{\delta \phi^2(x)}.
\end{align}

Recently, a more concrete relationship between AdS/CFT and stochastic quantization is proposed. In fact, a relationship between stochastic quantization and holographic Wilsonian renormalization group is made in \cite{Oh:2012bx}. In these papers authors suggest following identification among quantities appearing the both side.

\begin{enumerate}
    \item Let the imaginary time ``$t$'' in stochastic quantization and the radial coordinate ``$r$'' in AdS with quantum gravity theory be the same.
    \item Let the Euclidean action ${S_E}$ in stochastic quantization be $S_E=-2I_{os}$ for the holographic on-shell action $I_{os}$.
    \item The Fokker-Planck Hamiltonian $\mathcal{H}_{FP}(t)$ in stochastic quantization has the same form as the Hamiltonian $\mathcal{H}_{RG}(r)$ in the holographic renormalization group.
\end{enumerate}

\noindent We note that the authors in \cite{Oh:2012bx} prove that condition 3 can be derived from conditions 1 and 2 in \cite{Oh:2012bx}. They show that the following relationship holds when all three conditions are satisfied:

\begin{equation}
\label{1.8}
{<\phi_{p_1}(\epsilon)\phi_{p_2}(\epsilon)>}^{-1}_{H}|_{\epsilon=t}=<\phi_{p_1}(t)\phi_{p_2}(t)>^{-1}-\left.\frac{1}{2}\frac{\delta^2 S_E}{\delta\phi_{p_1}\delta\phi_{p_2}}\right|_{\phi=0},
\end{equation}

\noindent where $<\phi_{p_1}(t)\phi_{p_2}(t)>$ is a stochastic two-point correlation function and

\begin{equation}
{<\phi_{p_1}(\epsilon)\phi_{p_2}(\epsilon)>}^{-1}_{H}=
\left.\frac{\delta^2 S_B}{\delta\phi_{p_1}\delta\phi_{p_2}}\right|_{\phi=0}
\end{equation}

\noindent is a holographic double trace deformation.

This relation (\ref{1.8}) has been tested in various free theories:  the massless scalar theory in AdS$_2$ and the U(1) gauge fields theory in AdS$_4$ \cite{Oh:2012bx}, conformally coupled scalar theory in AdS$_d$ \cite{Jatkar:2013uga}, the massive scalar theory in AdS$_d$ \cite{Oh:2015xva}, the massless fermion theory in AdS$_d$ \cite{Oh:2013tsa}, and the massive fermion theory  in AdS$_d$ \cite{Moon:2017btx} are examined.

The authors in \cite{Oh:2021bxx} extend the relation into interacting theory. For the conformally coupled scalar in AdS$_6$, it is shown that the following relationship is hold:

\begin{align}
\label{1.10}
{<f_{p_1}(\epsilon)f_{p_2}(\epsilon)f_{p_3}(\epsilon)>}_{H}|_{\epsilon=t}=&
\frac{{<f_{p_1}(t)f_{p_2}(t)f_{p_3}(t)>}^{connected}}{<f_{p_1}(t)f_{-p_1}(t)><f_{p_2}(t)f_{-p_2}(t)><f_{p_3}(t)f_{-p_3}(t)>}
\\ \nonumber
&-\left.\frac{1}{2}\frac{\delta^3 S_E}{\delta f_{p_1}(t)\delta f_{p_2}(t)\delta f_{p_3}(t)}\right|_{f=0},
\end{align}

\noindent where ${<f_{p_1}(t)f_{p_2}(t)f_{p_3}(t)>}^{connected}$ is the connected stochastic three-point correlation function and

\begin{equation}
{<f_{p_1}(\epsilon)f_{p_2}(\epsilon)f_{p_3}(\epsilon)>}_{H}=
\left.\frac{\delta^3 S_B}{\delta f_{p_1}\delta f_{p_2}\delta f_{p_3}}\right|_{f=0}
\end{equation}

\noindent is a holographic triple trace deformation where the field $f$ is derived from the bulk field $\phi$, by a field redefinition as $f(x^{\mu})=r^{\frac{-d+1}{2}}\phi(x^{\mu})$.

In this paper, we extend the former relations (\ref{1.8}), (\ref{1.10}) to stochastic 4-point function and establish the relation between the stochastic 4-point function and holographic quadruple trace operator in order to obtain helpful results for extending the relation (\ref{1.8}) and establishing the relation between stochastic correlation function and holographic multiple trace operator in the theory with more general self-interaction. In the second chapter, a short introduction is made by redefining the field in the conformally coupled scalar theory in AdS$_{d+1}$, and the on-shell action in conformally coupled scalar theory in AdS$_4$ is obtained. In the third chapter, the holographic multiple trace deformation in conformally coupled scalar theory in AdS$_4$ is obtained, and in the fourth chapter, we obtain exact relation between stochastic 4-point function and quadruple trace deformation in conformally coupled scalar in AdS$_4$:
\begin{align}
\label{1.12}
\nonumber
{<f_{p_1}(\epsilon)f_{p_2}(\epsilon)f_{p_3}(\epsilon)f_{p_4}(\epsilon)>}_{H}&|_{\epsilon=t}
\\ \nonumber
=&-\frac{1}{<f_{p_1}(t)f_{-p_1}(t)><f_{p_2}(t)f_{-p_2}(t)><f_{p_3}(t)f_{-p_3}(t)><f_{p_4}(t)f_{-p_4}(t)>}
\\ \nonumber
&\times\Big(<f_{p_1}(t)f_{p_2}(t)><f_{p_3}(t)f_{p_4}(t)>+<f_{p_1}(t)f_{p_3}(t)><f_{p_2}(t)f_{p_4}(t)>
\\ \nonumber
&+<f_{p_1}(t)f_{p_4}(t)><f_{p_2}(t)f_{p_3}(t)>\left. -{<f_{p_1}(t)f_{p_2}(t)f_{p_3}(t)f_{p_4}(t)>}^{connected}\right)
\\
&-\left. \frac{1}{2}\frac{\delta^4 S_E}{\delta f_{p_1}(t)\delta f_{p_2}(t)\delta f_{p_3}(t)\delta f_{p_4}(t)}\right|_{f=0},
\end{align}

\noindent where ${<f_{p_1}(t)f_{p_2}(t)f_{p_3}(t)f_{p_4}(t)>}^{connected}$ is a connected stochastic 4-point correlation function, and

\begin{equation}
{<f_{p_1}(\epsilon)f_{p_2}(\epsilon)f_{p_3}(\epsilon)f_{p_4}(\epsilon)>}_{H}=
\left.\frac{\delta^4 S_B}{\delta f_{p_1}\delta f_{p_2}\delta f_{p_3}\delta f_{p_4}}\right|_{f=0}
\end{equation}

\noindent is a holographic quadruple trace deformation.

In the last chapter, a stochastic correlation function is obtained to test the extended relationship (\ref{1.12}), and it is substituted into (\ref{1.12}) to check that (\ref{1.12}) holds in the conformally coupled scalar theory AdS$_4$.

\section{A short introduction to conformally coupled scalar theory in AdS space}
\label{short}

To discuss the conformally coupled scalars theory in AdS$_{d+1}$ space, we consider an action,

\begin{equation}
\label{2.1}
   S=\int_{r>\epsilon} drd^{d}x\sqrt{g}\mathcal{L}(\phi,\partial\phi)+\int_{r=\epsilon} d^{d}x\mathcal{L}_{c.t.}(\phi,\partial\phi)+{S}_B,
\end{equation}

\noindent where the metric is given as

\begin{align}
\label{2.2}
ds^2= g_{\mu\nu}dx^\mu dx^\nu=\frac{1}{r^2}\left(dr^2+\sum_{i=1}^d dx^idx^i\right),
\end{align}

\noindent where $\phi$ is scalar field, the Greek indices as $\mu,\nu$ run from 1 to $d+1$ and so $x^{\mu}$ are $d+1$ the dimensional coordinate variable. $x^{d+1}=r$, so index $i$ run from 1 to d. $r=\epsilon$ is an arbitrary cut off in the radial direction. $x^i$ is the coordinate along the AdS boundary, $r$ is the radial coordinate in AdS space, $\mathcal{L}_{c.t.}$ is the counter term of the Lagrangian, and $S_B$ is the boundary action at $r=\epsilon$.

The detailed Lagrangian density of conformally coupled scalars theory is as follows:

\begin{equation}
   \mathcal{L}(\phi,\partial\phi)=\frac{1}{2}g^{\mu\nu}\partial_{\mu}\phi\partial_{\nu}\phi-\frac{1}{2}\frac{d^2-1}{4}\phi^2+\frac{\lambda}{4}\phi^{\frac{2(d+1)}{d-1}},
\end{equation}

\noindent $\lambda$ is self-interaction coupling constant. 
If we redefine field as

\begin{equation}
   \phi(x^{\mu})=r^{\frac{d-1}{2}}f(x^{\mu}),
\end{equation}

\noindent then the action of (\ref{2.1}) with the Lagrangian density of (\ref{2.2}) becomes

\begin{equation}
\label{2.5}
   S=\int_{r>\epsilon} drd^{d}x\left(\frac{1}{2}\partial_r f\partial_r f+\frac{1}{2}\partial_i f\partial_i f+\frac{\lambda}{4}\phi^{\frac{2(d+1)}{d-1}}\right)-\int_{r=\epsilon} d^{d}x\left(\mathcal{L}_{c.t.}+\frac{d-1}{4}\frac{f^2}{r}\right)+{S}_B(\epsilon).
\end{equation}

\noindent Since the term being proportional to $\frac{f^2}{r}$ is divergent on the AdS boundary, let us take $\mathcal{L}_{c.t.}=-\frac{d-1}{4}\sqrt{\gamma}\frac{f^2}{r}$ to cancel the $\frac{f^2}{r}$ term. $\gamma$ is the determinant of the boundary metric $\gamma_{ij}$ which is given by

\begin{equation}
   \gamma_{ij}=\frac{\partial x^{\mu}}{\partial x^{i}}\frac{\partial x^{\nu}}{\partial x^{j}}g_{\mu\nu}.
\end{equation}

\noindent The action (\ref{2.5}) can be effectively defined in upper half of $d+1$ dimensional flat space
\footnote{1-form gauge fields in AdS$_4$ can also be defined in effectively flat space. Studies on such theories are given in \cite{deHaro:2007eg,Jatkar:2012mm}}
, $\mathbb R^{d} \times [0,\infty)$. We note that stochastic quantization is also defined in such a product space where  $\mathbb R^{d}$ is where $S_E$ is defined and the stochastic time $t$ is lying along $[0,\infty)$.

We note that this theory has a $f^{\frac{2(d+1)}{(d-1)}}$ self-interaction. The exponent, $\frac{2(d+1)}{(d-1)}$ is usually fraction but it becomes an integer when $d=2,3,5$. When considering such theories, it would be reasonable to consider $d=3$ giving $f^4$-theory, $d=5$ giving $f^3$-theory, or $d=2$ giving $f^6$-theory. The $f^3$-theory is studied in \cite{Oh:2020zvm} and the $f^4$-theory is studied in \cite{Oh:2014nfa} in detail. In the future, I will briefly summarize the research in \cite{Oh:2014nfa} on the $f^4$-theory at $d=3$, which is the subject of this paper.

For $d=3$, the action of (\ref{2.5}) is given by

\begin{equation}
\label{2.7}
   S=\int_{r>\epsilon} drd^{3}x\left(\frac{1}{2}\partial_r f\partial_r f+\frac{1}{2}\partial_i f\partial_i f+\frac{\lambda}{4}\phi^{4}\right)+{S}_B(\epsilon).
\end{equation}

\noindent To expand this in boundary directional momentum space, we use the following Fourier transform,

\begin{equation}
\label{2.8}
  f(r,x)=\frac{1}{\sqrt{2\pi}^{3}}\int d^3p e^{-ip_ix_i}f_p(r),
\end{equation}

\noindent where $p_i$ is the momentum along the 3-dimensional AdS boundary directions. Then, the action in the momentum space is given as follows:

\begin{equation}
\label{2.9}
   S=\int_{r>\epsilon} drd^{3}p\left(\frac{1}{2}\partial_r f_p\partial_r f_{-p}+\frac{1}{2}p^2f_pf_{-p}+\frac{\lambda}{32\pi^3}\int d^{3}q d^{3}s d^{3}uf_pf_qf_sf_u \delta^{(3)}(p+q+s+u) \right)+{S}_B(\epsilon).
\end{equation}

\noindent Applying functional variation to the action (\ref{2.9}), the equation of motion in $f^4$-theory is obtained as follows:

\begin{equation}
\label{2.10}
   \partial^{2}_{r}f_p-p^2f_p-\frac{\lambda}{8\pi^3}\int d^{3}q d^{3}s d^{3}uf_pf_qf_s\delta^{(3)}(q+s+u-p)=0.
\end{equation}

\noindent We will solve this equation perturbatively up to order of $O(\lambda^1)$. A trial solution of the following form is substituted into (\ref{2.10}):

\begin{equation}
\label{2.11}
  f_p(r)=\bar f_p(r)+\lambda\tilde f_p(r)+\cdots.
\end{equation}

\noindent The equations of motion in $O(\lambda^0)$ and $O(\lambda^1)$ are given by,

\begin{equation}
   \partial^{2}_{r}\bar f_p-p^2\bar f_p=0\ (\lambda^0 order),
\end{equation}

\begin{equation}
   \partial^{2}_{r}\tilde f_p-p^2\tilde f_p-\frac{1}{8\pi^3}\int d^{3}q d^{3}s d^{3}u\bar f_q\bar f_s\bar f_u\delta^{(3)}(q+s+u-p)=0\ (\lambda^1 order).
\end{equation}

\noindent By solving the two equations, $\bar f_p (r)$ and $\tilde f_p (r)$ can be obtained as follows:

\begin{equation}
\label{2.14}
   \bar f_p(r)=\bar f_{0,p}e^{-|p|r},
\end{equation}

\begin{equation}
\label{2.15}
   \tilde f_p(r)=\tilde f_{0,p}e^{-|p|r}+\frac{1}{8\pi^3}\int d^{3}q d^{3}s d^{3}u\bar f_{0,q}\bar f_{0,s}\bar f_{0,u}\frac{e^{-(|q|+|s|+|u|)r}}{(|q|+|s|+|u|)^2-p^2}\delta^{(3)}(q+s+u-p),
\end{equation}

\noindent where $\bar f_{0,p}$ and $\tilde f_{0,p}$ are arbitrary boundary momentum dependent constants, so $\tilde f_{0,p}$ can be absorbed in $\bar f_{0,p}$ by setting $\tilde f_{0,p}$=0. Therefore, (\ref{2.11}) is given by

\begin{equation}
   f_p(r)=\bar f_{0,p}e^{-|p|r}+\frac{\lambda}{8\pi^3}\int d^{3}q d^{3}s d^{3}u\bar f_{0,q}\bar f_{0,s}\bar f_{0,u}\frac{e^{-(|q|+|s|+|u|)r}}{(|q|+|s|+|u|)^2-p^2}\delta^{(3)}(q+s+u-p)+\cdots.
\end{equation}

\noindent The solution expands near the AdS boundary as follows:

\begin{equation}
   f_p(r)=f_{p}^{(0)}-|p|f_{p}^{(0)}r-\frac{\lambda}{8\pi^3}r\int d^{3}q d^{3}s d^{3}uf_{q}^{(0)}f_{s}^{(0)}f_{u}^{(0)}\frac{(|q|+|s|+|u|-|p|)}{(|q|+|s|+|u|)^2-p^2}\delta^{(3)}(q+s+u-p)+O(r^2),
\end{equation}

\noindent $f_{p}^{(0)}$ is the value at $r=0$ of $f_p(r)$. More precisely, the $f_{p}^{(0)}$ is given by

\begin{equation}
f_{p}^{(0)}=\bar f_{0,p}+\frac{\lambda}{8\pi^3}\int d^{3}q d^{3}s d^{3}u\frac{\bar f_{0,q}\bar f_{0,s}\bar f_{0,u}}{(|q|+|s|+|u|)^2-p^2}\delta^{(3)}(q+s+u-p)+\cdots.
\end{equation}

By substituting the equation of motion (\ref{2.10}) into the action (\ref{2.9}), the following on-shell action $I_{os}$ can be obtained:

\begin{equation}
   I_{os}=\frac{1}{2}\int_{r=\epsilon} d^{3}p f_p(r)\partial_r f_p(r)-\frac{\lambda}{32\pi^3}\int drd^{3}pd^{3}q d^{3}s d^{3}u f_p(r)f_q(r)f_s(r)f_u(r)\delta^{(3)}(p+q+s+u)+S_B (\epsilon)+\cdots.
\end{equation}

\noindent Then, the on-shell action $I_{os}$ can also be obtained using various boundary conditions. Especially, in the Dirichlet boundary condition, it is not necessary to add boundary deformation. Then, substituting the boundary value of $f_p(r)$, the on-shell action $I_{os}$ in the Dirichlet boundary condition is as follows:

\begin{equation}
\label{2.19}
   I_{os}=-\frac{1}{2}\int_{r=\epsilon} d^{3}p |p|f_{p}^{(0)}f_{-p}^{(0)}-\frac{\lambda}{32\pi^3}\int d^{3}pd^{3}q d^{3}s d^{3}u \frac{f_{p}^{(0)}f_{q}^{(0)}f_{s}^{(0)}f_{u}^{(0)}}{|p|+|q|+|s|+|u|}\delta^{(3)}(p+q+s+u)+\cdots.
\end{equation}

\section{Description of the holographic Wilsonian renormalization group of conformally coupled $f^4$-theory}
\label{holographic Wilsonian renormalization group}

In this chapter, the holographic Hamilton-Jacobi equations and solutions of the conformally coupled $f^4$-theory are described. To derive the Hamiltonian of $f^4$-theory, we define the conjugate momentum of the field $f$. The Hamiltonian density is given by

\begin{equation}
   \mathcal{H}=\frac{1}{2}\Pi_p\Pi_{-p}-\frac{1}{2}p^2f_pf_{-p}-\frac{\lambda}{32\pi^3}\int d^{3}q d^{3}s d^{3}u f_{p}f_{q}f_{s}f_{u}\delta^{(3)}(p+q+s+u),
\end{equation}


\noindent where $\Pi_p=\partial_r f_{-p}$ is the conjugate momentum.

By using a fact that the variation of the total action $S$ with the cutoff $\epsilon$ is 0, and the conjugate momentum $\Pi_p$ is

\begin{equation}
   \Pi_p=\sqrt{g}\frac{\partial\mathcal{L}}{\partial(\partial_rf_p)}=\frac{\delta S_B}{\delta f_p},
\end{equation}

\noindent the Hamilton-Jacobi equation is derived as follows:

\begin{equation}
\label{3.3}
   \partial_\epsilon S_B=\int_{r=\epsilon}d^3p \left(-\frac{1}{2}\frac{\delta S_B}{\delta f_p}\frac{\delta S_B}{\delta f_p}+\frac{1}{2}p^2f_pf_{-p}+\frac{\lambda}{32\pi^3}\int d^{3}q d^{3}s d^{3}u f_pf_qf_sf_u\delta^{(3)}(p+q+s+u)\right).
\end{equation}

\noindent This equation describes the evolution of the boundary action $S_B$ as the radial cut-off $\epsilon$ changes. We will now solve this equation by introducing a trial solution given by

\begin{equation}
\label{3.4}
   S_B(\epsilon)=\Lambda(\epsilon)+\int d^3pJ_p(\epsilon)f_{-p}+\int d^3pD_{p}^{(2)}(\epsilon)f_{p}f_{-p}+\sum_{n=1}^{\infty}\int\prod_{i=1}^{n+2}d^3p_if_{p_i}D_{p_1,p_2,\cdots,p_{n+2}}^{(n+2)}(\epsilon)\delta^{(3)}\left(\sum_{n=1}^{n+2}p_i\right).
\end{equation}

Substituting this ansatz into the Hamilton-Jacobi equation, we get an expression in the form of series expansion order by order in weak field $f_p$. Demanding that the equation (\ref{3.3}) be identity equation in the field $f_p$, we get the following set of equations,

\begin{equation}
   \partial_\epsilon\Lambda(\epsilon)=-\frac{1}{2}\int_{r=\epsilon}d^3pJ_{p}(\epsilon)J_{-p}(\epsilon),
\end{equation}

\begin{equation}
   \partial_\epsilon J_{p}(\epsilon)=-2D_{-p}^{(2)}(\epsilon)J_p(\epsilon),
\end{equation}

\begin{equation}
   \partial_\epsilon D_{p}^{(2)}(\epsilon)=-3\int d^3p_1d^3p_2J_{-p}(\epsilon)D_{p_1,p_2,-p}^{(3)}(\epsilon)\delta^{(3)}(p_1+p_2-p)-2D_{-p}^{(2)}(\epsilon)D_{p}^{(2)}(\epsilon)+\frac{1}{2}p^2,
\end{equation}

\begin{equation}
   \partial_\epsilon D_{p_1,p_2,p_3}^{(3)}(\epsilon)=-4\int d^3p_1d^3p_2d^3p_3J_{-p}(\epsilon)D_{p_1,p_2,p_3,-p}^{(4)}(\epsilon)\delta^{(3)}(p_1+p_2+p_3-p)-2\sum^3_{n=1}D_{p_n}^{(2)}(\epsilon)D_{p_1,p_2,p_3}^{(3)}(\epsilon),
\end{equation}

\begin{align}
   \partial_\epsilon D_{p_1,p_2,p_3,p_4}^{(4)}(\epsilon)=&-5\int d^3p_1d^3p_2d^3p_3d^3p_4J_{-p}(\epsilon)D_{p_1,p_2,p_3,p_4,-p}^{(5)}(\epsilon)\delta^{(3)}(p_1+p_2+p_3+p_4-p)
   \\ \nonumber
   &-2\sum^4_{n=1}D_{p_n}^{(2)}(\epsilon)D_{p_1,p_2,p_3,p_4}^{(4)}(\epsilon)-\frac{9}{2}D_{p_1+p_2,p_3,p_4}^{(3)}(\epsilon)D_{p_1,p_2,p_3+p_4}^{(3)}(\epsilon)+\frac{\lambda}{32\pi^3},
\end{align}

\begin{align}
   \partial_\epsilon D_{p_1,p_2,\cdots,p_{n+2}}^{(n+2)}(\epsilon)=&-(n+3)\int d^3p_1dp_{n+2}J_{-p}(\epsilon)D_{p_1,p_2,\cdots,p_{n+2},-p}^{(n+3)}(\epsilon)\delta^{(3)}\left(\sum_{n=1}^{n+2}p_i-p\right)
   \\ \nonumber
   &-2\sum^{n+2}_{m=1}D_{p_m}^{(2)}(\epsilon)D_{p_1,p_2,\cdots,p_{n+2}}^{(n+2)}(\epsilon)
   \\ \nonumber
   &-\frac{1}{2}\sum^{n-1}_{m=1}(m+2)(n-m+2)D_{p_1,p_2,\cdots,p_{m+1},p_{m+2}+\cdots+p_{n+2}}^{(m+2)}(\epsilon)D_{p_1+p_2+\cdots+p_{m+1},p_{m+2},\cdots,p_{n+2}}^{(n-m+2)}(\epsilon).
\end{align}

The above set of equations is a collection of differential equations with the undetermined coefficients $D^{(n)}$ of $n$-multiple of the field $f_p$. This type of equation is difficult to solve, but if the source term $J_p$ is set to 0, the boundary cosmological constant $\Lambda$ becomes independent of $\epsilon$. In this simplest case $(J_p=0)$, $D^{(n)}$ is given as

\begin{equation}
   D_{p}^{(2)}(\epsilon)=\frac{1}{2}\frac{\partial_\epsilon f_p(\epsilon)}{f_p(\epsilon)},
\end{equation}

\begin{equation}
   D_{p_1,p_2,p_3}^{(3)}(\epsilon)=\frac{C_{p_1,p_2,p_3}^{(3)}}{f_{p_1}(\epsilon)f_{p_2}(\epsilon)f_{p_3}(\epsilon)},
\end{equation}

\begin{align}
   D_{p_1,p_2,p_3,p_4}^{(4)}(\epsilon)=&\int^\epsilon d\epsilon'\frac{f_{p_1}(\epsilon')f_{p_2}(\epsilon')f_{p_3}(\epsilon')f_{p_4}(\epsilon')}{f_{p_1}(\epsilon)f_{p_2}(\epsilon)f_{p_3}(\epsilon)f_{p_4}(\epsilon)}\left(\frac{\lambda}{32\pi^3}-\frac{9}{2}\frac {C_{p_1+p_2,p_3,p_4}^{(3)}}{f_{p_1}(\epsilon')f_{p_2}(\epsilon')f_{p_3+p_4}(\epsilon')}\frac{C_{p_1,p_2,p_3+p_4}^{(3)}}{f_{p_1+p_2}(\epsilon')f_{p_3}(\epsilon')f_{p_4}(\epsilon')} \right)
   \\ \nonumber   
   &+\frac{C_{p_1,p_2,p_3,p_4}^{(4)}}{\prod_{i=1}^4f_{p_i}(\epsilon)},
\end{align}

\begin{align}
   D_{p_1,p_2,\cdots,p_{n+2}}^{(n+2)}(\epsilon)=&-\frac{1}{2}\int^\epsilon d\epsilon'\frac{\prod_{i=1}^{n+2}f_{p_i}(\epsilon')}{\prod_{j=1}^{n+2}f_{p_j}(\epsilon)}
   \\ \nonumber
   &\times\left(\sum_{m=1}^{n-1}(m+2)(n-m+2)D_{p_1,p_2,\cdots,p_{m+1},p_{m+2}+\cdots+p_{n+2}}^{(m+2)}(\epsilon)D_{p_1+p_2+\cdots+p_{m+1},p_{m+2},\cdots,p_{n+2}}^{(n-m+2)}(\epsilon)\right)
   \\ \nonumber   
   &+\frac{C_{p_1,p_2,\cdots,p_{n+2}}^{(n+2)}}{\prod_{i=1}^{n+2}f_{p_i}(\epsilon)},
\end{align}

\noindent where $C^{(n)}$ is an integration constant and $f_p(\epsilon)$ is

\begin{equation}
\label{3.17}
   f_p(\epsilon)=C_psinh(|p|(\epsilon-\theta)),
\end{equation}

\noindent where $C_p$ and $\theta$ are arbitrary functions of momentum. We suppose $C^{(3)}=0$ for the boundary theory not to have triple trace deformation on it 
and that condition corresponds to the Dirichlet boundary condition on AdS boundary
\footnote{The authors consider a more general boundary condition on AdS boundary and its stochastic frame but they discuss the issues only in zero boundary momenta case\cite{Kim:2023bhp}.}
. In this case, $D_{p_1,p_2,p_3,p_4}^{(4)}(\epsilon)$ is given by

\begin{align}
\label{3.18}
   D_{p_1,p_2,p_3,p_4}^{(4)}(\epsilon)=&\frac{1}{\prod_{i=1}^4sinh(|p_i|(\epsilon-\theta))}\left(\frac{\lambda}{32{(2\pi)}^3}\left(\frac{sinh(\sum^4_{i=1}|p_i|(\epsilon-\theta))}{\sum^4_{i=1}|p_i|}\right.\right.
   \\ \nonumber
   &\left.-\sum^4_{i=1}\frac{sinh((\sum^4_{j=1}|p_j|-2|p_i|)(\epsilon-\theta))}{\sum^4_{j=1}|p_j|-2|p_i|}\left.+\sum^4_{i=2}\frac{sinh((|p_1|-\sum^4_{j=2}|p_j|+2|p_i|)(\epsilon-\theta))}{|p_1|-\sum^4_{j=2}|p_j|+2|p_i|}\right)\right)
   \\ \nonumber
   &+\frac{C_{p_1,p_2,p_3,p_4}^{(4)}}{\prod_{i=1}^4f_{p_i}(\epsilon)}.
\end{align}

\noindent In the above calculation, the following identity is used:

\begin{align}
   \prod_{i=1}^4sinh(|p_i|(\epsilon-\theta))=&\frac{1}{8}\left(cosh\left(\sum^4_{i=1}|p_i|(\epsilon-\theta)\right)\right.
   \\ \nonumber
   &-\sum^4_{i=1}cosh\left(\left(\sum^4_{j=1}|p_j|-2|p_i|\right)(\epsilon-\theta)\right)
   \\ \nonumber
   &\left.+\sum^4_{i=2}cosh\left(\left(|p_1|-\sum^4_{j=2}|p_j|+2|p_i|\right)(\epsilon-\theta)\right)\right).
\end{align}

\section{Derivation of exact relationship (1.14) between stochastic correlation function and holographic trace deformation}
\label{Derivation of exact relationship}

As mentioned in the introduction, the author in \cite{Oh:2021bxx} proposes conditions where they identify imaginary time ``$t$'' and Euclidean action $S_E$ in stochastic quantization with radial coordinates ``$r$'' and holographic on-shell action $I_{os}$ in AdS, respectively. When that condition is satisfied, they prove that the Fokker-Plank Hamiltonian $H_{FP}(t)$ in the stochastic quantization and the Hamiltonian $H_{RG}(t)$ in the holographic renormalization group can be set equal if there is no explicit ``$r$'' dependence in the action.

The two Hamiltonians appear equations (\ref{1.3}) and (\ref{1.7}) respectively. Since the two Hamiltonians can be set equal, the wave function $\phi_H(f,t)=e^{-S_B}, \phi_S(f,t)=P(f,t)e^{\frac{S_E}{2}}$ can also be set equal. To make it easy to compare the two wave functions, we redefine the stochastic probability distribution $P(f,t)$ as

\begin{equation}
   P(f,t)=e^{-S_P},
\end{equation}

\noindent where $S_P$ is the weight of correlation and has a following series form,

\begin{equation}
\label{4.2}
   S_P(f(t),t)=\sum_{i=2}^{\infty}\left[\prod_{j=1}^i\int d^3p_jf_{p_j}(t)\right]P_i(p_1,p_2\cdots p_i;t)\delta^{(3)}\left(\sum_{j=1}^ip_j\right),
\end{equation}

\noindent where $P_n$ is the coefficient of the $(f_p)^n$ term. We also expand Euclidean action $S_E$ into a following series form:

\begin{equation}
\label{4.3}
   S_E(f(t),t)=\sum_{i=2}^{\infty}\left[\prod_{j=1}^i\int d^3p_jf_{p_j}(t)\right]G_i(p_1,p_2\cdots p_i;t)\delta^{(3)}\left(\sum_{j=1}^ip_j\right).
\end{equation}

\noindent By comparing exponents of the two wave function forms, the following relationship can be derived:

\begin{equation}
   S_B=S_P-\frac{S_E}{2}.
\end{equation}

\noindent More precisely, by using $D^{(n)},P_n$ and $G_n$ given in (\ref{3.4}), (\ref{4.2}), and (\ref{4.3}) the relationship is given by

\begin{equation}
\label{4.4}
   D^{(n)}_{p_1,p_2,\cdots p_n}(\epsilon)|_{\epsilon=t}=P_n(p_1,p_2\cdots p_n;t)-\frac{1}{2}G_n(p_1,p_2\cdots p_n;t).
\end{equation}

Now, let us look at stochastic partition function.

\begin{equation}
   Z=\int[\mathcal{D}f_k]e^{-S_P}=\int[\mathcal{D}f_k]exp\left[-\sum_{n=2}\int P_n(k_n)\delta^{(3)}\left(\sum_{i=1}^nk_i\right)\prod^{n}_{j=1}f_jdf_j+\int d^3kJ_kf_k\right],
\end{equation}

\noindent where we consider that $P_n$ has order $O(\lambda^0)$ when $n = 2$. We suppose that $P_n$ for $n>2$, is put down by a small parameter. So $|P_2|>>|P_n|$ for $n>2$.

Now, let us consider the partition function up to $P_3$ term.
Then, the exponent of the partition function is expanded as

\begin{align}
   Z=&\int[\mathcal{D}f_k]\left(1-\int P_3(k_1,k_2,k_3)\delta^{(3)}(k_1+k_2+k_3)\prod^{3}_{j=1}\frac{\delta}{\delta J_{k_j}}d^3k_j+high\ order\right)
   \\ \nonumber
   &\times exp\left[-\frac{1}{4}\int d^3p_1d^3p_2 \frac{\delta^{(3)}(p_1+p_2)}{P_2(p_1,p_2)}J_{p_1}(t)J_{p_2}(t)\right].
\end{align}

\noindent From this form of the partition function, the stochastic two-point, connected three-point function can be obtained. They are respectively given by

\begin{equation}
   \left. <f_{p_1}(t)f_{p_2}(t)>=\frac{\delta}{\delta J_{p_1}}\frac{\delta}{\delta J_{p_2}}Z\right|_{J=0}=\frac{1}{2}\frac{\delta^{(3)}(p_1+p_2)}{P_2(p_1,p_2;t)}\ (up\ to\ \lambda^0),
\end{equation}

\begin{align}
   <f_{p_1}(t)f_{p_2}(t)f_{p_3}(t)>^{connected}=&\left. \frac{\delta}{\delta J_{p_1}}\frac{\delta}{\delta J_{p_2}}\frac{\delta}{\delta J_{p_3}}logZ\right|_{J=0}
   \\ \nonumber
   =&-3!P_3(p_1,p_2,p_3)<f_{p_1}(t)f_{-p_1}(t)><f_{p_2}(t)f_{-p_2}(t)><f_{p_3}(t)f_{-p_3}(t)>.
\end{align}

However, as we will see in the next chapter, the stochastic connected three-point function is zero for our case. Therefore, we demand $P_3$ vanish and we consider the next sub-leading, $P_4$. Then the distribution function is given as,

\begin{align}
   Z=&\int[\mathcal{D}f_k]\left(1-\int P_4(k_1,k_2,k_3,k_4)\delta^{(3)}(k_1+k_2+k_3+k_4)\prod^{4}_{j=1}\frac{\delta}{\delta J_{k_j}}d^3k_j+high\ order\right)
   \\ \nonumber
   &\times exp\left[-\frac{1}{4}\int d^3p_1d^3p_2 \frac{\delta^{(3)}(p_1+p_2)}{P_2(p_1,p_2)}J_{p_1}(t)J_{p_2}(t)\right].
\end{align}

\noindent The stochastic 4-point function can be obtained up to order $O(\lambda^1)$. It is given by

\begin{align}
\label{4.10}
   <f_{p_1}(t)&f_{p_2}(t)f_{p_3}(t)f_{p_4}(t)>^{connected}=\left. \frac{\delta}{\delta J_{p_1}}\frac{\delta}{\delta J_{p_2}}\frac{\delta}{\delta J_{p_3}}\frac{\delta}{\delta J_{p_4}}logZ\right|_{J=0}\ (up\ to\ \lambda^1)
   \\ \nonumber
&=-4!P_4(p_1,p_2,p_3,p_4,t)<f_{p_1}(t)f_{-p_1}(t)><f_{p_2}(t)f_{-p_2}(t)><f_{p_3}(t)f_{-p_3}(t)><f_{p_4}(t)f_{-p_4}(t)>.
\end{align}

\noindent $P_2$ can be obtained up to the order $O(\lambda^0)$, and $P_4$ up to the order $O(\lambda^1$). They are given by

\begin{equation}
   P_2(p_1,p_2)=\frac{1}{2}\frac{\delta^{(3)}(p_1+p_2)}{<f_{p_1}(t)f_{p_2}(t)>}\ (up\ to\ \lambda^0),
\end{equation}

\begin{equation}
    P_3(p_1,p_2,p_3)=-\frac{1}{3!}\frac{<f_{p_1}(t)f_{p_2}(t)f_{p_3}(t)>^{connected}}{<f_{p_1}(t)f_{-p_1}(t)><f_{p_2}(t)f_{-p_2}(t)><f_{p_3}(t)f_{-p_3}(t)>}\ (up\ to\ \lambda^1),
\end{equation}

\begin{align}
P_4(p_1,p_2,p_3,p_4)=-\frac{1}{4!}\frac{{<f_{p_1}(t)f_{p_2}(t)f_{p_3}(t)f_{p_4}(t)>}^{connected}}{<f_{p_1}(t)f_{-p_1}(t)><f_{p_2}(t)f_{-p_2}(t)><f_{p_3}(t)f_{-p_3}(t)><f_{p_4}(t)f_{-p_4}(t)>}\ (up\ to\ \lambda^1),
\end{align}

\noindent By using the following facts,

\begin{equation}
D^{(n)}_{p_1,p_2,\cdots p_n}(\epsilon)=\frac{1}{n!}\left.\frac{\delta^n S_B}{\delta f_{p_1}\delta f_{p_2}\cdots\delta f_{p_n}}\right|_{f=0},
\end{equation}

\begin{equation}
G_n(p_1,p_2\cdots p_n;t)=\frac{1}{n!}\left.\frac{\delta^n S_E}{\delta f_{p_1}\delta f_{p_2}\cdots\delta f_{p_n}}\right|_{f=0}.
\end{equation}

\noindent The relation (\ref{4.4}) becomes the followings when $n$=2 or 4 respectively:

\begin{equation}
{<f_{p_1}(\epsilon)f_{p_2}(\epsilon)>}^{-1}_{H}|_{\epsilon=t}=<f_{p_1}(t)f_{p_2}(t)>^{-1}-\left.\frac{1}{2}\frac{\delta^2 S_E}{\delta f_{p_1}\delta f_{p_2}}\right|_{f=0},
\end{equation}

\begin{align}
\label{4.18}
{<f_{p_1}(\epsilon)f_{p_2}(\epsilon)f_{p_3}(\epsilon)f_{p_4}(\epsilon)>}_{H}&|_{\epsilon=t}
\\ \nonumber
=&-\frac{{<f_{p_1}(t)f_{p_2}(t)f_{p_3}(t)f_{p_4}(t)>}^{connected}}{<f_{p_1}(t)f_{-p_1}(t)><f_{p_2}(t)f_{-p_2}(t)><f_{p_3}(t)f_{-p_3}(t)><f_{p_4}(t)f_{-p_4}(t)>}
\\ \nonumber
&-\left. \frac{1}{2}\frac{\delta^4 S_E}{\delta f_{p_1}(t)\delta f_{p_2}(t)\delta f_{p_3}(t)\delta f_{p_4}(t)}\right|_{f=0},
\end{align}

\noindent where 
the holographic quadruple trace deformation $<f( p_1 ) (\epsilon) f( p_2 ) (\epsilon) f ( p_3 ) (\epsilon) f( p_4 ) (\epsilon)>_H$ are as follows:


\begin{equation}
{<f_{p_1}(\epsilon)f_{p_2}(\epsilon)f_{p_3}(\epsilon)f_{p_4}(\epsilon)>}_{H}=
\left.\frac{\delta^4 S_B}{\delta f_{p_1}\delta f_{p_2}\delta f_{p_3}\delta f_{p_4}}\right|_{f=0}.
\end{equation}

\noindent We note that when $n$=3,

\begin{align}
\label{4.17}
{<f_{p_1}(\epsilon)f_{p_2}(\epsilon)f_{p_3}(\epsilon)>}_{H}|_{\epsilon=t}&=
\frac{{<f_{p_1}(t)f_{p_2}(t)f_{p_3}(t)>}^{connected}}{<f_{p_1}(t)f_{-p_1}(t)><f_{p_2}(t)f_{-p_2}(t)><f_{p_3}(t)f_{-p_3}(t)>}
\\ \nonumber
&=\left.\frac{\delta^3 S_E}{\delta f_{p_1}(t)\delta f_{p_2}(t)\delta f_{p_3}(t)}\right|_{f=0}=0,
\end{align}

\noindent trivially in our case.

\section{Stochastic correlation functions and check the relation (1.14)}
\label{check}

The Langevin equation in momentum space is

\begin{equation}
\label{5.1}
\frac{\partial f_p(t)}{\partial t}=-\frac{1}{2}\frac{\delta S_E}{\delta f_{-p}(t)}+\eta_p(t),
\end{equation}

\noindent where the Euclidean action $S_E$ is given by $S_E = -2I_{os}$ for the $I_{os}$ obtained in (\ref{2.19}), $\eta_p (t)$ is Gaussian noise with the following properties:

\begin{equation}
<\eta_p(t')\eta_{p'}(t'')>=\delta^{(3)}(p+p')\delta(t'-t''),
\end{equation}

\begin{equation}
\left<\prod^{2n-1}_{m=1}\eta_{p_m}(t_m)\right>=0,\ \textrm{and}  \ \left<\prod^{2n}_{m=1}\eta_{p_m}(t_m)\right>=\sum_{\textrm{all\ possible\ pairs\ of\ i\ and\ j}}\ \prod_{pairs}<\eta_{p_i}(t_i)\eta_{p_j}(t_j)>.
\end{equation}

Applying the Euclidean action $S_E$ of the $f^4$-theory to (\ref{5.1}), the Langevin equation is given by

\begin{equation}
\label{5.4}
\frac{\partial f_p(t)}{\partial t}=-|p|f_p(t)-\frac{\lambda}{(2\pi)^3}\int d^3qd^3sd^3uf_q(t)f_s(t)f_u(t)\frac{\delta^{(3)}(-p+q+u+s)}{|p|+|q|+|s|+|u|}+\eta_p(t)+O(\lambda^2).
\end{equation}

 We will solve this equation perturbatively up to order $O(\lambda^1)$. That is, the trial solution of the following form is substituted into (\ref{5.4}):

\begin{equation}
  f_p(t)=f_p^{(0)}(t)+\lambda f_p^{(1)}(t)+\cdots,
\end{equation}

\noindent Comparing the coefficients of $O(\lambda^0)$ and $O(\lambda^1)$ on both sides gives the following equations:

\begin{equation}
\frac{\partial f_p^{(0)}(t)}{\partial t}=-|p|f_p^{(0)}(t)+\eta_p(t)\ (\lambda^0 order),
\end{equation}

\begin{equation}
\frac{\partial f_p^{(1)}(t)}{\partial t}=-|p|f_p^{(1)}(t)-\frac{1}{(2\pi)^3}\int d^3qd^3sd^3uf_q^{(0)}(t)f_s^{(0)}(t)f_u^{(0)}(t)\frac{\delta^{(3)}(-p+q+u+s)}{|p|+|q|+|s|+|u|}\ (\lambda^1 order).
\end{equation}

\noindent By solving the two equations, $f_p^{(0)}(t)$ and $f_p^{(1)}(t)$ can be obtained as follows:

\begin{equation}
f_p^{(0)}(t)=\int^t_\tau dt'e^{-|p|(t-t')}\eta_p(t'),
\end{equation}

\begin{align}
f_p^{(1)}(t)=&-\frac{1}{(2\pi)^3}\int d^3qd^3sd^3u\int^t_\tau dt'\int^{t'}_\tau dt''dt'''dt''''e^{-|p|t-(|q|+|s|+|u|-|p|)t'+|q|t''+|s|t'''+|u|t''''}
\\ \nonumber
&\times\eta_q(t'')\eta_s(t''')\eta_u(t'''')\frac{\delta^{(3)}(-p+q+u+s)}{|p|+|q|+|s|+|u|}.
\end{align}

\noindent Using these solutions, stochastic n-point correlation function up to order of $\lambda^0$ or $\lambda^1$ can be obtained. We note that the correlation functions are obtained up to tree level (we loop contributions). One point function is

\begin{equation}
<f_p(t)>=\int^t_\tau dt'e^{-|p|(t-t')}<\eta_p(t')>=0\ (up\ to\ \lambda^0).
\end{equation}

\noindent Tree level two point function is given by

\begin{align}
<f_p(t)f_{p'}(t)>=&\int^t_\tau dt'e^{-|p|(t-t')}\int^t_\tau dt''e^{-|p'|(t-t'')}<\eta_p(t')\eta_{p'}(t'')>
\\ \nonumber
=&-\frac{sinh(|p|(\tau-t))}{|p|e^{-|p|(\tau-t)}}\delta^{(3)}(p+p')\ (up\ to\ \lambda^0),
\end{align}

\noindent and tree level three point function vanish as

\begin{align}
\nonumber <f_p(t)f_{p'}(t)f_{p''}(t)>=&\int^t_\tau dt'dt''dt'''e^{-(|p|+|p'|+|p''|)t+|p|t'+|p'|t''+|p''|t'''}<\eta_p(t')\eta_{p'}(t'')\eta_{p''}(t''')>
\\ \nonumber
&-\frac{\lambda}{(2\pi)^3}\int^t_\tau dt'dt''dt'''e^{-(|p|+|p'|+|p''|)t+|p|t'+|p'|t''+|p''|t'''}
\\ \nonumber
&\times \int d^3qd^3sd^3u\int^{t'}_\tau dt''''dt'''''dt''''''e^{-(|q|+|s|+|u|)t'+|q|t''''+|s|t''''''+|u|t'''''''}
\\ \nonumber
&\times\frac{\delta^{(3)}(-p+q+u+s)}{|p|+|q|+|s|+|u|}<\eta_q(t'''')\eta_s(t'''''')\eta_u(t'''''''')\eta_{p'}(t'')\eta_{p''}(t''')>
\\ \nonumber
&+(p\leftrightarrow p')+(p\leftrightarrow p'')
\\
=&0\ (up\ to\ \lambda^1).
\end{align}

\noindent Our tree level four point function has two different contributions. One is disconnected part which is given by

\begin{align}
\nonumber <f_{p_1}(t)f_{p_2}(t)f_{p_3}(t)f_{p_4}(t)>=&\int^t_\tau dt' e^{-|p_1|(t-t')}\int^t_\tau dt''e^{-|p_2|(t-t'')} \int^t_\tau dt''' e^{-|p_3|(t-t''')}\int^t_\tau dt'''' e^{-|p_4|(t-t'''')}
\\ \nonumber
&\times <\eta_{p_1}(t')\eta_{p_2}(t'')\eta_{p_3}(t''')\eta_{p_4}(t'''')>
\\ \nonumber
=&\int^t_\tau dt' e^{-|p_1|(t-t')}\int^t_\tau dt''e^{-|p_2|(t-t'')} \int^t_\tau dt''' e^{-|p_3|(t-t''')}\int^t_\tau dt'''' e^{-|p_4|(t-t'''')}
\\ \nonumber
&\times (<\eta_{p_1}(t')\eta_{p_2}(t'')><\eta_{p_3}(t''')\eta_{p_4}(t'''')>
\\ \nonumber
&+<\eta_{p_1}(t')\eta_{p_3}(t''')><\eta_{p_2}(t'')\eta_{p_4}(t'''')>
\\ \nonumber
&+<\eta_{p_1}(t')\eta_{p_4}(t'''')><\eta_{p_2}(t'')\eta_{p_3}(t''')>)
\\ \nonumber
=&<f_{p_1}(t)f_{p_2}(t)><f_{p_3}(t)f_{p_4}(t)>+<f_{p_1}(t)f_{p_3}(t)><f_{p_2}(t)f_{p_4}(t)>
\\
&+<f_{p_1}(t)f_{p_4}(t)><f_{p_2}(t)f_{p_3}(t)>\ (\lambda^0\ order),
\end{align}

\noindent However, our relation (\ref{4.18}) is only about connected part, which is given by

\begin{align}
\nonumber <f_{p_1}(t)f_{p_2}(t)f_{p_3}(t)&f_{p_4}(t)>=<f_{p_1}^{(0)}(t)f_{p_2}^{(0)}(t)f_{p_3}^{(0)}(t)f_{p_4}^{(1)}(t)>+(p_1\leftrightarrow p_4)+(p_2\leftrightarrow p_4)+(p_3\leftrightarrow p_4)
\\ \nonumber
=&-\frac{\lambda}{(2\pi)^3}\int^t_\tau dt' e^{-|p_1|(t-t')}\int^t_\tau dt''e^{-|p_2|(t-t'')} \int^t_\tau dt''' e^{-|p_3|(t-t''')}\int^t_\tau dt'''' e^{-|p_4|(t-t'''')}
\\ \nonumber
&\times\int d^3k_1d^3k_2d^3k_3\int^{t''''}_\tau dt_1 e^{-|k_1|(t''''-t_1)}\int^{t''''}_\tau dt_2e^{-|k_2|(t''''-t_2)} \int^{t''''}_\tau dt_3 e^{-|k_3|(t''''-t_3)}
\\ \nonumber
&\times \frac{\delta^{(3)}(-p_4+k_1+k_2+k_3)}{|p_4|+|k_1|+|k_2|+|k_3|}(<\eta_{p_1}(t')\eta_{p_2}(t'')\eta_{p_3}(t''')\eta_{k_1}(t_1)\eta_{k_2}(t_2)\eta_{k_3}(t_3)>
\\
&+(p_1\leftrightarrow p_4,t'\leftrightarrow t'''')+(p_2\leftrightarrow p_4,t''\leftrightarrow t'''')+(p_3\leftrightarrow p_4,t'''\leftrightarrow t'''')\ (\lambda^1\ order).
\end{align}

\noindent The final form of the connected diagram are

\begin{align}
\nonumber <f_{p_1}(t)f_{p_2}(t)&f_{p_3}(t)f_{p_4}(t)>^{connected\ \lambda^1\ order}
\\ \nonumber
=&-\frac{3\lambda}{4(2\pi)^3}e^{\left(\sum^4_{i=1}|p_i|\right)(\tau-t))}\Bigg(e^{-\left(\sum^4_{i=1}|p_i|\right)(\tau-t))}\Bigg.
\\ \nonumber
&-\sum^4_{i=1}\frac{(\sum^4_{j=1}|p_j|-|p_i|)e^{-\left(\sum^4_{j=1}|p_j|-|p_i|\right)(\tau-t)}-|p_i|e^{\left(\sum^4_{j=1}|p_j|-|p_i|\right)(\tau-t))}}{\sum^4_{j=1}|p_j|-2|p_i|}
\\ \nonumber
&\Bigg.-\sum^4_{i=2}\frac{(\sum^4_{j=2}|p_j|-|p_i|)e^{-\left(|p_1|-\sum^4_{j=2}|p_j|+2|p_i|\right)(\tau-t)}-(|p_1|+|p_i|)e^{\left(|p_1|-\sum^4_{j=2}|p_j|+2|p_i|\right)(\tau-t))}}{|p_1|-\sum^4_{j=2}|p_j|+2|p_i|}\Bigg)
\\
&\times\frac{1}{\prod^4_{i=1}|p_i|}\frac{\delta^{(3)}(\sum^4_{i=1}|p_i|)}{\sum^4_{j=1}|p_j|}.
\end{align}

\noindent The equation (\ref{4.17}) for the relationship of the stochastic 3-point correlation function is valid when $C_{(p_1,p_2,p_3)}^{(3)}=0$.

Finally, the right side of (\ref{1.12}) is calculated to confirm the relationship to the stochastic 4-point correlation function:

\begin{align}
\label{5.17}
R.H.S\ of\ (1.14)=&\frac{3\lambda}{4(2\pi)^3}\frac{1}{\prod_{i=1}^4sinh(|p_i|(t-\tau))}\left(\frac{sinh(\sum^4_{i=1}|p_i|(t-\tau))}{\sum^4_{i=1}|p_i|}\right.
\\ \nonumber
 &\left.\left.-\sum^4_{i=1}\frac{sinh((\sum^4_{j=1}|p_j|-2|p_i|)(t-\tau))}{\sum^4_{j=1}|p_j|-2|p_i|}+\sum^4_{i=2}\frac{sinh((|p_1|-\sum^4_{j=2}|p_j|+2|p_i|)(t-\tau))}{|p_1|-\sum^4_{j=2}|p_j|+2|p_i|}\right)\right).
\end{align}

\noindent Equation (\ref{5.17}) exactly corresponds to the quadruple trace operator in (\ref{3.18}) when $t=r,\tau=\theta,C_{(p_1,p_2,p_3,p_4)}^{(4)}=0$.

\section*{Acknowledgement}
J.H.O thanks his W.J. and Y.J. and he thanks God. This work was supported by the National Research Foundation of Korea(NRF) grant funded by the Korea government(MSIT) (No.2021R1F1A1047930).

\end{document}